\documentclass[sigconf]{acmart}
\usepackage{subfigure}
\usepackage{listings}
\usepackage{xcolor}
\newcommand{\etal}{\textit{et al.}}

\AtBeginDocument{%
  \providecommand\BibTeX{{%
    \normalfont B\kern-0.5em{\scshape i\kern-0.25em b}\kern-0.8em\TeX}}}

\setcopyright{acmcopyright}
\copyrightyear{2022}
\acmYear{2022}
\acmDOI{10.1145/1122445.1122456}

\acmConference[ESEC/FSE 2022]{The 30th ACM Joint European Software Engineering Conference and Symposium on the Foundations of Software Engineering}{14 - 18 November, 2022}{Singapore}

\begin{document}

%\title{Neural Graph Attention Networks for the Estimation of \linebreak Program Termination}
%\title{Estimating and Understanding Program Termination with Graph Neural Networks}
\title{Using Graph Neural Networks for Program Termination}

\author{Yoav Alon}
\email{yoav.alon@bristol.ac.uk}
\orcid{0000-0003-1432-3057}
\affiliation{
  \institution{University of Bristol}  
  \city{Bristol}  
  \country{UK}
  \postcode{BS8 1TH}
}

\author{Cristina David}
\email{cristina.david@bristol.ac.uk}
\orcid{0000-0002-9106-934X}
\affiliation{
  \institution{University of Bristol}  
  \city{Bristol}  
  \country{UK}
  \postcode{BS8 1TH}
}

\renewcommand{\shortauthors}{Alon and David}

\begin{abstract}
  
Termination analyses investigate the termination behavior of programs,
intending to detect nontermination, which is known to
cause a variety of program bugs (e.g. hanging programs,
denial-of-service vulnerabilities).  Beyond formal approaches,
various attempts have been made to estimate the termination behavior
of programs using neural networks.  However, the majority of these
approaches continue to rely on formal methods to provide
strong soundness guarantees and consequently suffer from similar
limitations.
%Consequently, they also suffer from limitations specific to formal methods.
In this paper, we move away
from formal methods and embrace the stochastic nature of machine
learning models. Instead of aiming for rigorous guarantees
that can be interpreted by solvers, our objective is to provide an
estimation of a program's termination behavior and of the likely reason for
nontermination (when applicable) that a programmer can use for
debugging purposes.  Compared to previous approaches using neural
networks for program termination, we also take advantage of the graph
representation of programs by employing Graph Neural Networks.  To
further assist programmers in understanding and debugging
nontermination bugs, we adapt the notions of attention and semantic
segmentation, previously used for other application domains, to
programs.  Overall, we designed and implemented classifiers for
program termination based on Graph Convolutional Networks and Graph
Attention Networks, as well as a semantic segmentation Graph Neural
Network that localizes AST nodes likely to cause nontermination. We also
illustrated how the information provided by semantic segmentation can
be combined with program slicing to further aid debugging.
\end{abstract}




\maketitle

\section{Introduction} \label{sec:intro}

%Intro termination problem 
Termination analysis describes a classical decision problem in computability theory where program termination has to be determined. It is critical for many applications such as software testing, where nonterminating programs will lead to infinite executions.
As proved by Turing in 1936, a general algorithm that solves the termination problem for all possible program-input pairs doesn't exist \cite{TuringHalting}.
%Classical approaches 
While there are a large number of works on termination analysis, the majority of them employ formal symbolic reasoning~\cite{DBLP:conf/esop/DavidKL15,DBLP:conf/tacas/CookSZ13, DBLP:conf/pldi/CookPR06, DBLP:conf/popl/GuptaHMRX08, DBLP:conf/pldi/ChatterjeeG0Z21, ClassicalTermination}. 
%Neural estimation attempts
In recent years, various attempts have been made to estimate termination behavior using neural networks. For instance, Giacobbe \etal \cite{NeuralTermination} introduced an approach where neural networks are trained as ranking functions (i.e. monotone maps from the program's state space to
well-ordered sets). A similar idea is employed in \cite{LearningProb}, where Abate et al. use a neural network to fit ranking supermartingale (RMS) over execution traces.
Given that program analysis tasks such as termination analysis are generally expected to provide formal guarantees, these works use 
satisfiability modulo theories (SMT) solvers to show the validity of their results. 
While promising, they still face limitations specific to formal symbolic methods. Namely, programs need to be translated to a symbolic representation to generate the verification conditions that are then passed to the solver. Additionally, these verification conditions may be expressed in undecidable logical fragments or may require extra program invariants for the proof to succeed.

In this paper, we move away from formal methods and lean into the stochastic nature of machine learning models. Instead of looking for rigorous formal guarantees that can be interpreted by solvers, {\em our objective is to provide an estimation of a program's termination behavior, as well as localizing the likely cause of nontermination
  % innermost infinitely looping constructs
  %and of the reason for nontermination
  (when applicable)} that a programmer can use for debugging purposes.
Our work also serves as a study of the applicability of machine learning techniques previously used for other classes of applications to program analysis.
In particular, as explained next, we use Graph Neural Networks (GNNs)~\cite{GNNs} and Graph Attention Networks (GANs) \cite{GraphAttention}. % and semantic segmentation \cite{SemanticGnn}.

Instead of looking at execution traces like the aforementioned works, we are interested in using the source code with
the assumption that it contains patterns that can assist in understanding its functionality.
Notably, program analysis techniques generally work on source code, and specifically on graph representations of programs.
To emulate this for machine learning, we make use of GNNs, which 
are a class of neural networks optimized to perform various analyses on graph-structured data.
GNNs are gaining a lot of interest as they are being used to analyze
graph-based systems denoting social networks~\cite{DBLP:conf/aaai/WuLXWC20}, physical systems~\cite{sanchez2018graph}, knowledge graphs~\cite{hamaguchi2017knowledge}, point-cloud classification~\cite{GNNPointCloudClassification} etc.
Additionally, GNNs have recently been applied to program analysis tasks such as variable misuse detection and type inference~\cite{GNNProgramAnalysis}, and self-supervised bug detection and repair \cite{DBLP:journals/corr/abs-2105-12787}. 

Inspired by~\cite{GNNProgramAnalysis,DBLP:journals/corr/abs-2105-12787}, we use GNNs to estimate program termination.
Our baseline program termination classifier is based on the Graph Convolutional Networks (GCN)~\cite{GraphConvLayer}.

On its own, estimating a program's termination behavior doesn't provide a lot of practical
help a programmer interested in understanding and debugging a nontermination bug.
Rather, we would like to provide additional information such as the code location 
corresponding to the likely root cause of the failure (in our case nontermination).
This objective is similar to that of fault localization, which takes as input a set of
failing and passing test cases, and produces a ranked list of potential causes of failure~\cite{DBLP:conf/issta/LeLGG16}.

As opposed to fault localization techniques, we are interested in investigating using the
mechanisms of attention and semantic segmentation from machine learning.
To the best of our knowledge, we are the first ones to use attention and segmentation
in the context of programs. %also gnn's for termination. [more important] 

Attention is a technique that mimics cognitive attention.
Intuitively, it enhances some parts of the input data while diminishing others with the expectation that the network is focusing on a small, but important part of the data.
In our context, we use attention to get an intuition about the instructions relevant for the estimation of the termination behavior. This allows us to visualize
those parts of the program that the neural network focuses on to estimate its termination behavior.
To integrate attention in our work, we build another program termination classifier inspired by the Graph Attention Network (GAT) architecture described in~\cite{GraphAttention}. Given the varied influence that different instructions in a program have on its termination behavior, we also expect the attention mechanism to improve the results of classification when compared to the GCN-based baseline.

To localize the likely cause of the nontermination behavior, we use semantic segmentation.
Usually used in image recognition, the goal of semantic
image segmentation is to label each pixel of an image with a
particular class, allowing one to identify objects belonging to that
class (e.g. given an image, one can identify a cat that appears in it). In our work, we use the same principle for programs to identify those statements that cause nontermination vs those that don't.

To further aid debugging, we also show how to use the information provided by
semantic segmentation to carve out a nonterminating slice from the
original program (i.e. a smaller subprogram exhibiting the same
nontermination behavior as the original program). Intuitively,
such a smaller program is easier to understand and debug.

Our experimental evaluation for multiple datasets, both custom and based on benchmarks from software verification competitions, confirms a high ability to generalize learned models to unknown programs. 

The main contributions of this research are as follows:  

\begin{itemize}
  \item We designed and implemented a GCN-based architecture for the binary classification of program termination. 
  \item We designed and implemented a GAT-based architecture that improves the termination classification using a self-attention mechanism and allows visualization of the nodes relevant when estimating termination. 
  \item We designed and implemented a semantic segmentation GAT that localizes nodes causing nontermination. In particular, in this work, we try to localize the outermost infinitely looping constructs. We illustrate how the information provided by semantic segmentation can be combined with program slicing to further aid debugging.
  \item We devised datasets for both classification and segmentation of program termination. 
  %Finding datasets for training machine learning-based program analyses is very challenging given that, despite a large amount of available code, the majority of it is not labeled.
\end{itemize}

\section{Preliminaries on Graph Neural Networks} \label{sec:gnns}
 
GNNs are an effort to apply deep learning to non-euclidean data represented as graphs. These networks have recently gained a lot of interest as they are being used to analyze
graph-based systems~\cite{DBLP:conf/aaai/WuLXWC20,sanchez2018graph,hamaguchi2017knowledge,GNNPointCloudClassification}.
%denoting social networks~\cite{DBLP:conf/aaai/WuLXWC20}, physical systems~\cite{sanchez2018graph}, knowledge graphs~\cite{hamaguchi2017knowledge}, point-cloud classification~\cite{GNNPointCloudClassification} etc.
A comprehensive description of existing approaches for GNNs and their applications can be found in \cite{GNNs,GNNSurvey2,ZHOU202057}.
 
The basic idea behind most GNN architectures is graph convolution or message passing, which is adapted from Convolutional Neural Networks (CNN).
Each vertex (node) in the graph has a set of attributes, which we refer to as a feature vector or an embedding.
Then, a graph convolution estimates the features of a graph node in the next layer as a function of the neighbors' features.
By stacking GNN layers together, a node can eventually incorporate information from other nodes further away.
For instance, after $n$ layers, a node has information about the nodes $n$ steps away from it.
 
%% The necessity to use graph convolution instead of regular fully connected neural networks on the adjacency matrix of a graph is based on multiple factors. First, for the processing of graphs, permutation invariance is required. Isomorph graphs may appear different but are structurally identical. For that reason, adjacency matrices can not be used with feed-forward networks for graph data, which is sensitive to changes in node order. Furthermore, the graph structure is non-euclidean, a reason why machine learning around graphs is commonly called geometric deep learning. A representation or embedding of data is learned, where the structure of nodes is conserved and node embeddings are generated.
 
Given a graph $G = (V, E)$, where $V$ denotes the set of vertices and $E$ represents the set of edges, message passing works as follows.
For each node $i \in V$ and its embedding $h_i^{(l)}$ at time step $l$, the embeddings $h_j^{(l)}$ of its neighbours $j \in N(i)$  are aggregated and the current node's embedding is updated to $h_i^{(l+1)}$
 using aggregation function $A$ and update function $U$: 
 \begin{equation}
    h_i^{(l+1)} = U^{(l)}(h_i^{l}, A^{(l)}( \{ h_j^{(l)}, \forall j \in N(i)\} ))
 \end{equation}

We use $N(i)$ to describe the set of direct neighbors of $i$. 
Each GNN architecture varies the implementation of the update $U$ and aggregation $A$ functions used for message passing.
In this paper, we make use of two GNN architectures: Graph Convolutional Networks (GCNs) and Graph Attention Networks (GATs), which we will discuss in subsequent sections.

\begin{figure*}[]
  \centering
  
  \subfigure[]{\includegraphics[height=6cm]{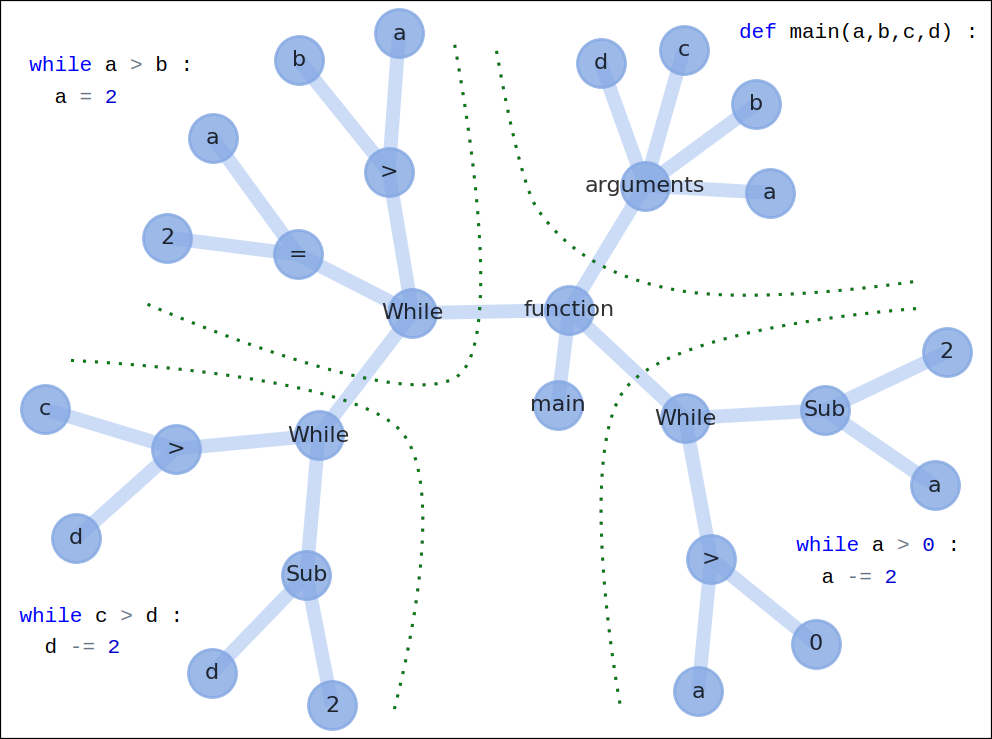}}\quad
  \subfigure[]{\includegraphics[height=6cm]{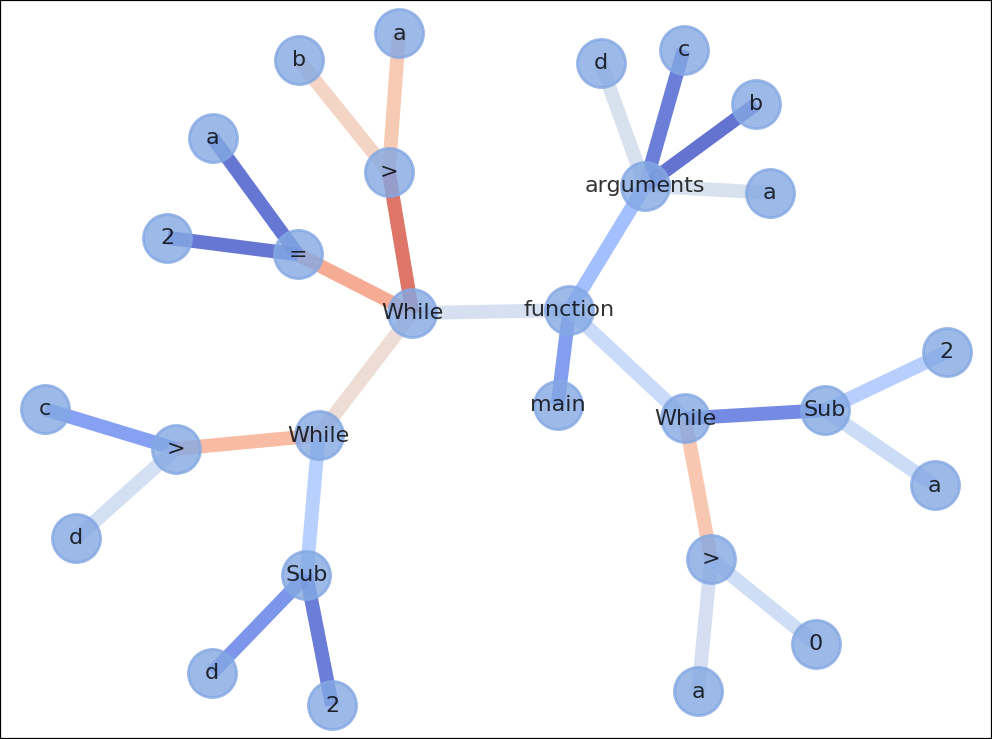}}\quad
  \subfigure[]{\includegraphics[height=6cm]{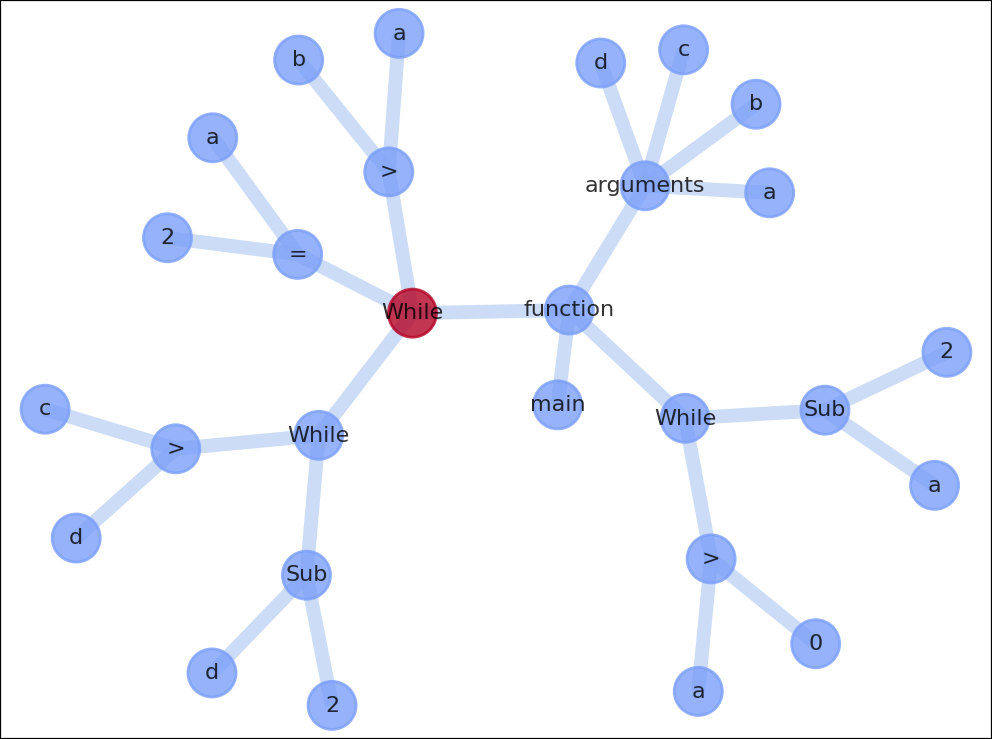}}\quad
  \subfigure[]{\includegraphics[height=6cm]{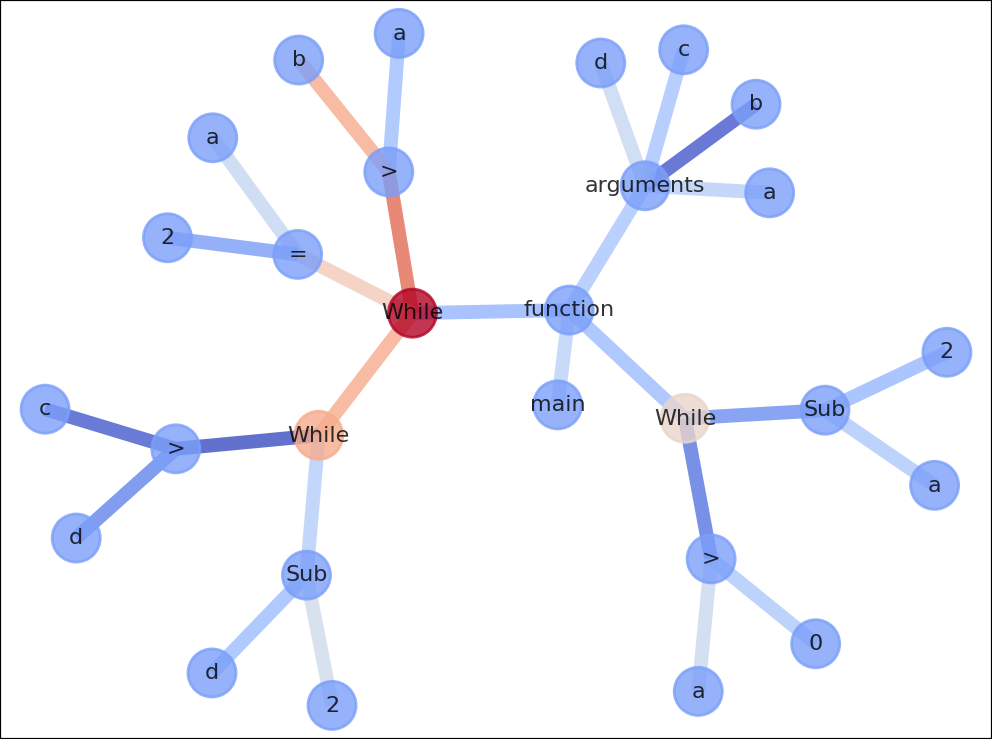}}\quad
  \caption{(a) AST (for clarity, we point out the corresponding lines of code). (b) Visualization of attention for binary classification as edge color from blue for low attention to red for high attention. The edges with high attention are those connecting the \texttt{while} nodes to their loop guards. Among these, the highest attention is given to the nonterminating outer loop. (c) Result of semantic segmentation, where the node corresponding to the outermost infinite loop is coloured in red. (d) Semantic segmentation visualized together with the attention extracted from the segmentation network. Best viewed in color. }
  \label{ExampleGraphs}
\end{figure*}

\section{Description of our technique}\label{sec:formal}

In this section, we provide details about our technique.
At a high level, we first convert programs to feature graphs and then feed them into a GNN.
We will describe each step, including the structure of the models for termination classification and semantic segmentation of the nodes responsible for nontermination.

\subsection{Generation of feature graphs} \label{sec:conversion}
% graph level task
% supervised setting

There are many graph representations of a program. In this work, we 
start from the Abstract Syntax Tree (AST), which is a homogenous, undirected graph, where each node denotes a construct occurring in the program.
We picked ASTs as the starting point because they are simple to understand and construct while containing all the necessary information for investigating a program's termination.
As future work, we may consider different graph representations of programs.

We start by generating the AST of each program in our datasets.
Then, all ASTs are converted into {\em feature graphs} by converting each node to a local feature vector in local representation or 'one-hot'-encoding.
For this purpose, all the nodes in the ASTs generated for each dataset are gathered in a dictionary. The dictionary holds each encoding as key with the instruction as value.

\paragraph{Running example}
%% \todo{Alternative running example description:}
%% For illustration, we refer to our running example in Figure~\ref{fig:alternative-running-example}. This example is constructed by putting together three benchmarks from the termination category of the SV-COMP 2022 competition \cite{sv-comp, SVCOMP} (the first conditional and loop correspond to Cairo.c, the second to Cairo\_step2-3.c, and the last to Cairo\_step2-1.c from the termination-crafted subcategory). While the first and the last loops terminate, the middle one is nonterminating.
%% In particular the nontermination of the middle loop is not easy for symbolic tools to prove. For instance, the winner of the SV-COMP 2021 termination category, {\sc UAutomizer}~\cite{DBLP:conf/cav/HeizmannHP13} fails to determine this fact. 

For illustration, we refer to our running example in Figure~\ref{fig:running-example}, which contains three loops.
Out of them, the first outer loop is nonterminating for any initial value of
$b$ less than 2 and of $a$ greater than $b$. Note that for machine integers (i.e. bit-vectors), the inner loop does terminate even when starting from a $c$ greater than $d$ because $d$
will eventually underflow, thus triggering a wrap-around behavior.

%% For an initial $c$ greater than $d$, the inner loop of the first two nested loops will never terminate if we consider mathematical integers, where values don't underflow.

\begin{figure}[]
  \centering
  
   \subfigure{\includegraphics[height=2.7cm]{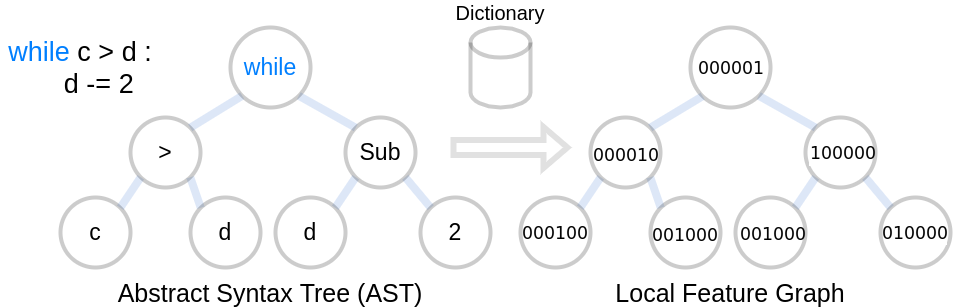}}\quad

   \caption{Conversion of an AST to a local feature graph (using 'one-hot'-encoding). } 
   \label{fig:FeatureGraph}
 \end{figure}

The running example is first converted to its AST as shown in Figure~\ref{ExampleGraphs}(a). Then, the AST is converted to the initial feature graph (before any convolutions).
For reasons of space, we don't show the whole feature graph, but rather just the subgraph corresponding to the instructions \texttt{while c>d: d-=2}
in Figure~\ref{fig:FeatureGraph}. The dictionary maps each node in the AST to a one-hot encoding. For instance, the \texttt{while} node is mapped to $0 \cdots 00010000$. Similarly,
both nodes corresponding to variable \texttt{d} have the same encoding, $0 \cdots 00000010$. As we use the same dictionary for a whole dataset, identical nodes in the dataset will
have the same encoding.

\begin{figure}
  \begin{lstlisting}
 def main(a,b,c,d) :
 
   while a > b:
     a = 2 
     while c > d:
       d -= 2

   while a > 0: 
     a -= 2

\end{lstlisting}
\caption{Running example}
\label{fig:running-example}
\end{figure}

  %% def test(a,b,c,d) : 
  %%     while a > b :
  %%         a -= 1 
  %%         while c > d:  
  %%             d -= 2 
  %%     return True

%% \begin{figure}
%% \begin{lstlisting}
%%   int foo(int x, int y, int z) {
%%     if (x > 0) {
%%       while (x != 0) 
%%         x = x - 1;
%%    }

%%    if (y > 0) {
%%      while (y != 0)
%%        y = y - 2;
%%    }
	
%%    if (z > 0) {
%%      while (z != 0 && z!= -1) 
%%        z = z - 2;
%%    }
%%    return 0;
%%  }
%% \end{lstlisting}
%% \caption{Alternative running example}
%% \label{fig:alternative-running-example}
%% \end{figure}

\subsection{Binary classification of termination based on GCN}\label{sec:classification}

For our first attempt at a binary classifier for program termination, we make use of GCN. In particular, we take inspiration from the architecture in \cite{GraphConvLayer} in a supervised training setting. Currently, we do not train on recursive programs, where the nontermination could be caused by infinite recursion. We plan to do this as future work.
For now, our datasets contain programs with potentially infinite loops. 

Our termination classification is a graph level task, where we predict an attribute for the entire graph (in our case the termination behavior of the corresponding program)
by aggregating all node features.  
%We use label 1 for terminating programs and label 0 for non-terminating ones.  

As explained in Section~\ref{sec:gnns}, GNNs use message passing to aggregate the information about a node’s neighbors to update the node's value.
According to~\cite{GraphConvLayer}, we use:
%% For the classification of termination, we obtain the new node embedding for the current node $c$ at time step $k{+}1$ with the following formula, where we use the aggregator operator from \cite{GraphConvLayer}:
 
\begin{equation}\label{eq:gcn}
    h_i^{(l+1)} = \sigma(b^{(l)} + \sum_{j \in N(i)} \frac{1}{c_{ij}} h_j^{(l)}W^{(l)})
\end{equation}

where $c_{ij} = \sqrt{|N(i)|} \sqrt{|N(j)|}$, $N(i)$ describes the set of direct neighbours of $i$, and $\sigma$ is the activation function. 

%Based on a better training performance with fast convergence, the
Best results are achieved using a rectified linear unit (ReLu) activation function. The weights $W^{(l)}$ are initialized using Glorot uniform and the bias $b$ with zero. The experiments of~\cite{GraphConvLayer} with other aggregation operators rather than the one used here, such as Multi-Layer-Perceptron aggregator \cite{GraphMLP} and Graph Attention Networks \cite{GraphAttention}, achieved similar results. 

%Specific to our method 
For the architecture in \cite{GraphConvLayer}, the graph convolution layers generate node-wise features. In our architecture, given that termination classification is a graph-level task, we choose to mean all the nodes once graph convolution is done and pass the resulting mean feature vector through three fully connected layers. Then, we apply a softmax function on the resulting two-dimensional vector to achieve a binary estimation. The resulting vector is then optimized by computing the cross-entropy loss with the ground truth.

\paragraph{Running example} For our running example in Figure~\ref{fig:running-example}, our classifier correctly concludes that the program is nonterminating. As aforementioned, we operate in a supervised training setting, meaning that our classifier has already been trained on the required dataset. We will give more details on the training and testing in Section~\ref{sec:Experiments}. % and we now may deploy them. 

\subsection{Binary classification of termination based on GAT}

%Say something about GAT
If we revisit the update rule of GCNs given by Equation~\ref{eq:gcn},  it uses the coefficient $\frac{1}{c_{ij}} = \frac{1}{\sqrt{|N(i)|} \sqrt{|N(j)|}}$. Intuitively, this coefficient suggests the importance of node $j$'s features for node $i$ and it is heavily dependent on the structure of the graph (i.e. each node's neighbours).
The main idea of GATs~\cite{GraphAttention} is to compute this coefficient implicitly rather than explicitly as GCNs do by considering it to be a learnable attention mechanism. %, instead of solely reliant on the graph's structure.
In the rest of the section, we refer to this coefficient as the {\em attention score}.

The main observation that led us to use GATs is the fact that not all instructions in a program are equally important when investigating its termination behavior.
For instance, loops tend to be more important than straight-line code, and should therefore be given increased attention. Thus, our task naturally lends itself to using the attention mechanism.
%
%% Recent progress in self-attention mechanisms have led to the development of graph attention layers \cite{GraphAttention} that apply normalized attention scores to every source node as part of the aggregation. In such a way, better performance for most benchmark tasks is achieved, and additionally, the importance of relational patterns that influence classification can be extracted. 
%
Moreover, our intention for this work is not only to design a program termination classifier but also to gain insights into its decisions. Using GATs helps us in this direction as it allows us to visualize those nodes that influence the decision related to termination. 

%For a single graph attention layer
Next, we explain next how the aggregation and update of node features $h_i^{(l)}$ to $h_i^{(l+1)}$ for iteration $l$ is derived.
Initially, each node feature is passed through a simple linear layer by multiplying the feature with a learnable weight matrix $W^{(l)}$:  

\begin{equation}
z_i^{(l)} =W^{(l)}h_i^{(l)}
\end{equation}
To obtain the pair-wise importance $e_{ij}^{(l)}$ of two neighbor features $h_i^{(l)}$ and $h_j^{(l)}$ we concatenate the previously computed embeddings $z_i^{(l)}$ and $z_j^{(l)}$ to $(z_i^{(l)}|z_j^{(l)})$. Then, the concatenated embeddings are fed into another linear layer with weight matrix $W_2^{(l)}$ that learns the attention scale $a^{(l)}$. Additionally, a leakyReLU activation function \cite{LeakyRelu} is applied to introduce non-linearity:
\begin{equation}
a^{(l)} =  \text{LeakyReLU}(W^{(l)} (z_i^{(l)}|z_j^{(l)}))
\end{equation}
After computation of $a$, corresponding to the activation scale, the pair-wise importance $e_{ij}^{(l)}$ of two neighbor features is: 

\begin{equation}
    e_{ij}^{(l)} =\text{LeakyReLU}(\vec a^{(l)^T}(z_i^{(l)}|z_j^{(l)}))
\end{equation}
As noted by \cite{GraphAttention}, the resulting attention scale can be considered as edge data. In our problem setting, this gives us insight into the importance of two connected syntax tree nodes with respect to the program's termination. 

To normalize the attention scores for all incoming edges we a apply a softmax layer:
\begin{equation}
\alpha_{ij}^{(l)} =\frac{\exp(e_{ij}^{(l)})}{\sum_{k\in N(i)}^{}\exp(e_{ik}^{(l)})}    
\end{equation}
Then, the neighbor embeddings are aggregated and scaled by the final attention scores $\alpha_{ij}^{(l)}$:
\begin{equation}
    h_i^{(l+1)} =\sigma\left(\sum_{j\in N(i)} {\alpha^{(l)}_{ij} z^{(l)}_j }\right)
\end{equation}
%With a neighbor relation of i and j, here denoted as  $j\in \mathcal{N}(i)$. Thus updating features based on neighbor features weighted by attention assigned to them. 

\begin{figure}[]
  \centering
  
   \subfigure{\includegraphics[height=4cm]{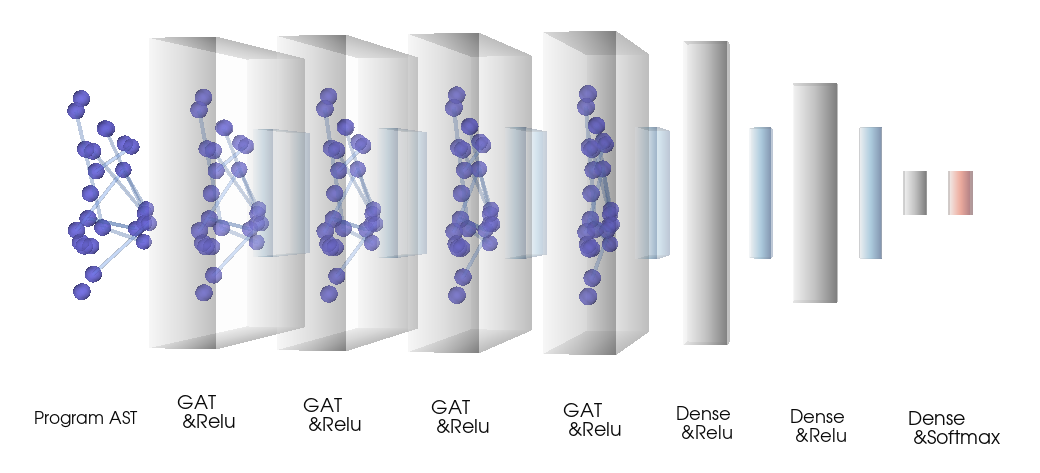}}\quad

   \caption{GAT-based architecture for the binary classification of program termination.} 
   \label{fig:Classification}
 \end{figure}

The architecture of our GAT-based termination classifier is shown in Figure~\ref{fig:Classification} and includes multiple connected GAT layers with ReLu activation. Graph convolutions extract features to an intermediate representation, denoted by the last GAT layer with ReLu activation. Conversely to regular graph attention networks, we devise a prediction on the program's termination by extracting the mean of node features and using dense layers with a final softmax layer. The attention information is extracted from the last GAT layer and can be associated with the AST edges. 

\paragraph{Running example}
Similar to the GCN, the GAT-based classifier identifies that the running example is nonterminating.
Additionally, in Figure~\ref{ExampleGraphs}(b) we also obtain a visualization of edge-wise attention from blue for low attention to red for high attention. Notably, the edges with high attention are those connecting the \texttt{while} nodes to their loop guards. The extraction of attention gives us an intuition about which nodes have a high influence on the final prediction provided by the classifier. %In this case, as expected, they are the loops.
Although for this example, the edge connecting the nonterminating outer loop with its loop guard gets the highest attention, the difference between the attention scores linked to the three loops is not reliable enough to differentiate between the three loops and determine the cause for nontermination.

\subsection{Semantic segmentation of nodes causing nontermination}

While GATs provide some insight into the decision made by the classifier (by allowing us to visualize those AST nodes influencing the decision),
we want to find the likely cause of nontermination.
In particular, in this work, we try to localize
the outermost infinitely looping constructs.
Note that in the case where we have several nested loops and the outer one is infinitely looping, all of the inner ones will also be visited infinitely often.
In such a situation, we identify the outer loop as the likely cause of nontermination, i.e. the outermost infinitely visited loop. As aforementioned, for now, we do not train on recursive programs, where the nontermination could be caused by infinite recursion. Thus, in our setting nontermination can only be caused by infinite loops.

%exact statements responsible for nontermination whenever a program is nonterminating.
Identifying the outermost infinitely visited loop is not possible with the information that we gained up until this point. Although in Figure~\ref{ExampleGraphs}(b), the edge between the nonterminating outer loop and its guard gets the highest attention, the edges connecting the other two loops and their guards also have high attention scores. Thus, it is hard to differentiate between the different loops based on the attention score alone.

%% two \texttt{while} loops has the highest attention,
%% but the inner loop is the one that iterates infinitely. Based on the attention score alone, there is no way of differentiating between the two loops.

To achieve our goal of identifying the outermost infinitely visited loop, we make use of semantic segmentation. %~\cite{}.
Semantic segmentation has been primarily used in image recognition to label each pixel of an image with a particular class, allowing one to identify objects belonging to that class.
%For instance, it can be used to identify all the pixels forming a cat. 
%
Sometimes in the literature, the semantic segmentation of graphs is also named 'node-wise classification'  or 'node-classification'. 

In this work, we attempt to extend the same principle that was applied in image recognition to programs by using semantic segmentation to identify those statements that cause nontermination. If we consider that program statements can belong to two classes, namely those that cause nontermination and those that don't, we can envision that segmentation may help identify the former.

The graph network must convert feature vectors to an estimation. For this purpose, it is trained on the loss between ground truth and prediction on a node level. This can be accomplished with most segmentation loss functions such as a simple cross-entropy loss for binary segmentation with ground truth $y$ and prediction $\hat{y}$:
\begin{equation}
L(y, \hat{y}) = -(y \log (\hat{y}) + (1 - y) \log (1 - \hat{y}))
\end{equation}

Although the above loss enables the setup of an initial training session, it is not ideal to eradicate hard negatives. 
To improve and continue training beyond binary cross-entropy convergence, we use focal loss \cite{LossFunctions, FocalLoss}, an extension of cross-entropy, which  down-weights simple samples and gives additional weight to hard negatives: 
\begin{equation}
FL(p_t) = - \alpha_t (1 - p_t)^{\gamma} log(p_t)
\end{equation}
with $\gamma, \alpha \in [0,1]$ and the modulation fact $(1-p_t)^{\gamma}$, where
$p_t = p$ for $y = 1$ and $p_t = 1-p$ otherwise.

Following the same message passing method explained in Section~\ref{sec:classification}, 
the new node features hold the segmentation prediction.

 \begin{figure}[]
   \centering
  
   \subfigure{\includegraphics[height=4.5cm]{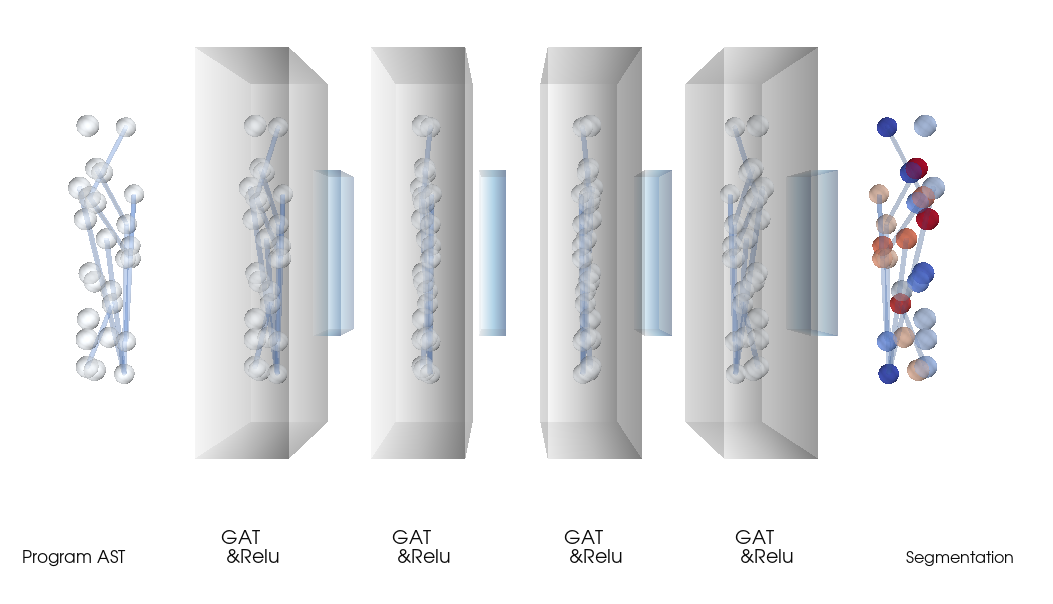}}\quad
    
   \caption{Proposed GAT-based architecture for the segmentation of nodes causing nontermination. %% Graph attention convolution generates a semantic segmentation resulting in a node-wise estimation of likelihood to cause non-termination. By the attention mechanism, edge-wise scores denote the weight of features based on relational patterns.
   }
   \label{fig:Segmentation}
 \end{figure}

 Our architecture for semantic segmentation is visualized in Figure \ref{fig:Segmentation}. Graph attention convolution generates a semantic segmentation resulting in a node-wise estimation of likelihood to cause nontermination. By the attention mechanism, edge-wise scores denote the weight of features based on relational patterns.

\paragraph{Running example}
 The result of semantic segmentation for our running example is highlighted in Figure~\ref{ExampleGraphs} (c), where the red-colored node denotes the outermost infinite loop.
 Figure~\ref{ExampleGraphs} (d) shows the result of segmentation with attention.
 
\subsection{Using the result of segmentation for debugging}\label{sec:discussion-debugging}

One issue with the result of segmentation is that it only highlights the node corresponding to the head of the outermost nonterminating loop, rather than all the statements contributing to
nontermination. This is obvious for our running example in Figure~\ref{ExampleGraphs} (c), where segmentation only identifies the node corresponding to the nonterminating \texttt{while} loop.

This was a conscious decision on our part, meant to simplify the construction of the required training datasets.
While there are a huge number of programs available online, it's not at all straightforward to use them for training machine learning-based techniques for program analysis. The main reason for this is that such programs are not labeled with the result of any analysis, and manually labeling them is difficult and error-prone. In particular, for the termination classification task,
we must know
% to build training (and testing) datasets for the termination classification task, we must know
which programs are terminating and which are not.
Even more problematic, for semantic segmentation, the annotation should, in principle, identify all the instructions contributing to nontermination.
For instance, for our running example, it should identify lines 3 and 4. %% (note that the outer loop also needs to be included to ensure that the inner one is reachable)
In general, it can be very difficult to annotate such datasets.

In this work, to simplify the annotation for segmentation, we only label the node corresponding to the head of the outermost infinite loop as the reason for nontermination.
More details on the way we generate our datasets are given in Section~\ref{sec:Experiments}.

While identifying the outermost infinite loop is already useful, debugging can be further aided by finding the rest of the instructions that contribute to nontermination.
This can be achieved with slicing~\cite{10.1145/358557.358577}.
%
%% These instructions can be obtained by slicing the program according to a slicing criterion that preserves the values of the variables used in the guard of the infinite
%% loop (i.e. those variables will have the same values in the original and sliced programs).
%
In general, slicing is a program analysis technique that aims to extract parts of a program according to a particular slicing criterion 
(e.g. it can extract instructions responsible for the write accesses to a particular variable).
%in our case the values of the variables used by the guard of a potentially infinite loop).
In our scenario, given a nonterminating loop returned by segmentation, we can use several slicing criteria in order to isolate the faulty loop, while preserving the nontermination behavior highlighted by segmentation. One option, which we illustrate below on the running example in Figure~\ref{fig:running-example}, is a criterion aiming to preserve the reachability of a control flow point (in particular, we are interested in the control flow point at the entrance to the nonterminating loop). Among other tools, such a slicing capability is provided by the software analysis platform Frama-C~\cite{DBLP:journals/cacm/BaudinBBCKKMPPS21}. Other slicing criteria can be used, e.g. preserving the values of the variables used by the guard of the infinite loop.
%(i.e. those variables will have the same values in the original and sliced programs). 
Slicing has been previously used in the context of debugging nontermination, e.g. failure-slices in~\cite{DBLP:conf/ppdp/NeumerkelM99}.

Going back to our running example in Figure~\ref{fig:running-example},
%where lines 5,6, 9, and 10 have been added. Similar to the running example,
once segmentation has identified the first outer loop as the likely culprit for nontermination,
we slice the program such that we preserve the reachability of the control flow point at the entrance to the body of the nonterminating outer loop.
This gives us the program in Figure~\ref{fig:sliced-running-example}.
This sliced program preserves the nontermination behavior of the original program (induced by the loop identified by segmentation) while cutting down the syntactic structure, which corresponds to reducing its state space.

Intuitively, cutting down the state space of the program makes it more amenable to other existing
debugging techniques such as fuzzing. For illustration, we have asked libFuzzer~\cite{libfuzzer} (a library for coverage-guided fuzz testing) to search for an input that triggers execution of the sliced program longer than 5s. LibFuzzer was able to find the input $a$ = 165017090 and 
$b$ = -183891446 within 6s. This input triggers the nontermination behavior and can assist in further debugging. Conversely, when asking libFuzzer to find such input for the original in Figure~\ref{fig:running-example},
it required 12s. We expect the difference between the two running times to increase for larger programs.

Overall, the debugging workflow that we envision follows the following steps:
(1) our classifier estimates that a program is nonterminating;
(2) our technique based on GNNs and semantic segmentation finds the likely
root cause of nontermination; (3) slicing cuts down the syntactic structure of the program
and implicitly its state space, while aiming to maintain the  faulty nontermination behavior;
(4) fuzzing finds faulty inputs for the sliced program.

If a program contains several nontermination bugs, the debugging workflow above may need to be repeated until all of the bugs are eliminated.

%% While for such a small program, there isn't a considerable difference between the time taken by fuzzing the original and the slide programs, this will have a considerable impact when debugging
%% larger programs.

%% . Next,
%% slicing such that the values of \texttt{c} and \texttt{d} (from the inner loop's guard) are preserved will return a program where
%% lines 5, 6, as well as the additional loop at lines 9 and 10, have been removed. These extra lines have no involvement in the program's nonterminating behavior and shouldn't be considered during debugging.

%% This notion of a potentially nonterminating slice resulting slice is similar to that of failure-slices, which are an executable reduced fragment of a logical program. Similar to our scenario,
%% failure-slices expose those parts of the program that may cause non-termination~\cite{}.

%\begin{figure}
%\begin{lstlisting}
%  def test(a,b,c,d,e,f) :
%    while a > b:
%      a -= 1 
%      while c > d:
%        e = d + f
%        f = f - c
%        d -= 2

%    while e < f: 
%      e = e + d

%     return True
%\end{lstlisting}
%\caption{Modified running example}
%\label{fig:modified-running-example}
%\end{figure}

\begin{figure}
\begin{lstlisting}
 def main(a,b) :
   while a > b:
     a = 2 

\end{lstlisting}
\caption{Nonterminating slice of the running example}
\label{fig:sliced-running-example}
\end{figure}

\section{Experiments}
\label{sec:Experiments}

\subsection{System}
For our experiments, we used the deep learning framework PyTorch \cite{PyTorch} with the GNN package DGL (Deep Graph Library) and various packages for preprocessing and visualization. The experiments were performed using an RTX 2080 8GB GPU on a local machine.
The architecture of the GNNs used, depending on the particular experiment consists of between 4 and 6 graph layers. While this is typical for the effective training of GNNs, it could be extended using residual blocks \cite{DeepGCN}.

\subsection{Binary classification of program termination}

We start by discussing our experimental evaluation for the binary classification of program termination. Given a program, our classifier extracts a termination estimation, i.e. whether the program is terminating or nonterminating. Concerning the latter, a program is considered nonterminating if there exists an input for which the program's execution is infinite.

\subsubsection{Classification datasets}

Our experiments on classification and semantic segmentation of the cause for nontermination require different datasets. This is because classification data requires program-level annotations (a program is labeled as terminating or nonterminating), whereas segmentation requires node-wise annotations (each node in the AST is either a cause for nontermination or not). In this section, we discuss the datasets used for classification, whereas more details on those used for segmentation can be found in Section~\ref{sec:datasets-segmentation}.
 
For classification, we use a total of four datasets, out of which two are based on existing benchmarks and two are generated by us. Concerning the datasets containing existing benchmarks, the first one, DS-SV-COMP, contains C programs from the termination category of the SV-COMP 2022 competition \cite{sv-comp, SVCOMP}. In particular, this dataset contains all programs from the following subcategories: termination-crafted, termination-crafted-lit, termination-numeric, and termination-restricted-15. 
In total, there are 55 nonterminating programs and 194 terminating. For a total of 249 programs, we count a total of 386 loops with a maximum of five nested loops for the program \texttt{NO\_04.c}. The second dataset of existing benchmarks, DS-TERM-COMP, contains 150 C programs from the Termination Competition \cite{TERMCOMP} with a total of 452 loops and a maximum of 2 nested loops per program. These benchmarks are selected such that there is no overlap with those in the DS-SV-COMP dataset. %Both datasets feature nested loops. 

The benchmarks in both DS-SV-COMP and DS-TERM-COMP expect inputs meaning that their termination behavior depends on non-deterministic values. These benchmarks come already labeled as terminating or nonterminating, where the nontermination label indicates that there exists an input for which the program's execution doesn't terminate.
The pre-processing of the datasets includes the generation of the AST representation for each program and the conversion to a feature graph using DGL.

For efficiency reasons, we consider batches of 30 graphs using the $dgl{.}batch$.
The two datasets have an associated dictionary featuring all the distinct AST nodes from all the programs.
Intuitively, this dictionary gives us the vocabulary used by the benchmarks.
For more details on the generation of feature graphs, see Section~\ref{sec:conversion}.

As previously discussed in Section~\ref{sec:discussion-debugging}, one of the main challenges of our work is the fact that, while there are many available programs, only a few are already labeled as terminating or nonterminating  (as we could see above, even existing software verification competitions only consider a relatively small number of benchmarks).
However, machine learning techniques generally require large training data, which is not available in our setting. 
Due to this, we chose to also generate our additional custom datasets. For the first custom dataset, DS1, we used the dictionary generated for DS-SV-COMP and DS-TERM-COMP, meaning that the vocabulary of DS1 is the same as that of DS-SV-COMP and DS-TERM-COMP. This is important as it allows us to train on DS1 and only use DS-SV-COMP and DS-TERM-COMP for testing. DS1 contains 950 C programs. The second custom dataset, DS2, contains 950 generated Python programs, out of which 800 are used for training and 150 for testing. 

To label the benchmarks in  DS1 and DS2, we fuzz test for nontermination by generating a large number of inputs for each program. % depending on the number of input variables, for which we execute the corresponding program. 
If the execution time reaches a predefined timeout for at least one of the inputs, then we label the program as nonterminating. The dataset is balanced by providing an equal number of terminating and nonterminating samples.

Apart from DS-SV-COMP and DS-TERM-COMP, each dataset 
is split into training and test samples by the general rule of approximately 80/20 depending on the dataset size. The exact split for each set is specified in Table~\ref{tab:datasets}.

\begin{table*}
  \centering 
  \begin{tabular}{l c c c c c c } 
  \textbf{Dataset} & Origin & Language & Experiment & Training samples & Test samples \\ % & Instructions  \\ 
  \midrule   
  \textbf{DS-SV-COMP} & SV-Comp & C & classification & - & 249 \\ %& 122/148 \\   
  \textbf{DS-TERM-COMP} & TermComp & C & classification & - & 150 \\ %& 118/148 \\   
  \midrule   
  \textbf{DS1} & custom & C & classification & 800 & 150 \\ %& 148 \\   
  \textbf{DS2} & custom & Python & classification & 800 & 150 \\ %& 84 \\   
  \midrule   
  \textbf{DS-Seg-Py1} & custom & Python & semantic segmentation & 180 & 50 \\ % & 89 \\     
  \midrule   
  \textbf{DS-Seg-C} & custom & C & semantic segmentation & 180 & 50 \\ % & 89 \\   
  
  \bottomrule 
  \end{tabular}
  \caption{Datasets used in this work.}    
  \label{tab:datasets} 
  \end{table*}

\subsubsection{Training}

As explained in Section~\ref{sec:formal}, we deploy two different neural networks, one based on GCN \cite{GraphConvLayer} and another one based on GAT \cite{GraphAttention}. In our experiments, we found that a number of 4 layers was sufficient for both the GCN and GAT architectures, respectively. Training is performed using an Adam-optimizer with an initial learning rate of 0.0001. We use a regular cross-entropy loss function and record various metrics such as the  Receiver Operating Characteristic (ROC) curve and the Precision-Recall during training both for validation and testing. 

For a significant evaluation, we perform a total of ten training sessions until convergence of the test metrics. %and extract the following ROC and Precision-Recall metrics. %% Macro-average will compute the ROC for each class and compute the average, while micro-average aggregates the contributions of classes for metric computation.  

\subsubsection{Results and evaluation}

%[paragraph zero, about metrics ]  =====================================
To judge the performance of our classifier, we use the Precision-Recall (PR) and Receiver Operating Characteristic (ROC) and their respective Area Under the Curve (AUC) and Average Precision (AP). AUC takes values from 0 to 1, where value 0 indicates a perfectly inaccurate test and 1 reflects a perfectly accurate test. 

PR is used as an indication for the tradeoff between precision and recall for different thresholds. Consequently, high recall and high precision reflect in a high area under the curve. High precision is based on a low rate of false positives and a high recall is based on a low rate of false negatives. 

%The true positive rate, also called recall, is the fraction of correctly predicted 1’s among all true 1’s.
%The precision = TP / (FP + TP) is the fraction of correctly predicted 1’s among all predicted 1’s.

The ROC curve plots the true positive rate versus the false positive rate for each threshold. The closer to the top-left corner a ROC curve is, the better, and the diagonal line corresponds to random guessing. A high area under the curve indicates a better classification performance. %A diagonal reference line denotes the distance to a random result.

%[first paragraph (without differentiating between positive and negative classes)]
The results of the experimental evaluation of our classifier are summarized in Table~\ref{tab:clasResult}, where we record the AUC for positive (terminating) instances, the AUC for negative (nonterminating) instances, and the mean Average Precision (mAP), which is calculated as the average of the AP for terminating and nonterminating instances.

The first and arguably most important observation is that, 
with values of over 0.82 for all mAPs for both PR and ROC, we can conclude that classification results are significant and that we achieve a high ability to generalize for unknown data. 

%[second paragraph GCN vs. GAT]  =====================================
The second remark refers to the comparison of GCN and GAT. As observed in Table~\ref{tab:clasResult}, the PR and ROC mAP numbers are generally higher for GAT than GCN, suggesting that the application of graph attention improves the result of the binary classification (we use bold font for the higher mAP numbers). This indicates that the applied self-attention mechanism does assign a higher weight to patterns responsible for deciding the termination behavior.

%[third paragraph (with respect to classes)] =====================================
The third comparison we are interested in is between the classification of terminating and nonterminating programs. 
We notice that the PR AUC negative tends to be higher than AUC positive, meaning that the classifier performs better at identifying nonterminating programs. We conjecture that nonterminating patterns are easier to identify based on relational probabilistic patterns that are likely to cause nontermination.

\begin{figure}[]
  \centering

  \includegraphics[height=6.5cm]{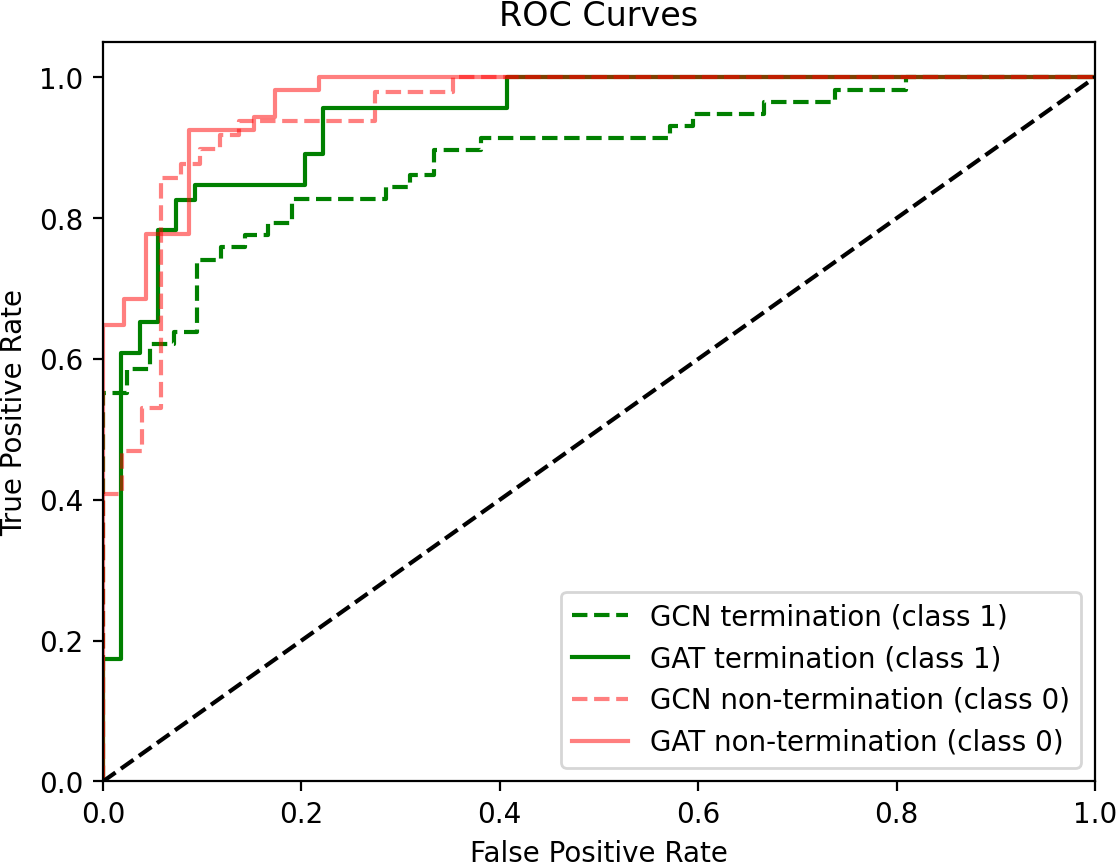}
  %\subfigure[]{\includegraphics[height=6cm]{images/new_roc.png}}\quad  
  %\subfigure[]{\includegraphics[height=6cm]{images/new_pr.png}}\quad

  \caption{ROC curves per class (class 0 corresponds to nonterminating programs and class 1 to terminating ones) and experimental configuration.    
  } 
  
  \label{fig:binResults}
\end{figure}

In Figure~\ref{fig:binResults} we visualize the ROC curves per class and layer type. The better performance for nontermination classes is reflected in higher red curves while the better performance of GAT layers is represented by higher continuous lines compared to doted CGN curves. 

%
%We believe that the difference in the performance of for negative and positive classes is represented better by the PR than the ROC metrics due to the imbalanced class data for DS-SV-COMP and DS-TERM-COMP. 

%ROC robustness to threshold/ confidence - With respect to PR -> better in negative samples (non-terminating programs)
%PR robustness in terms of types of mistakes - - With respect to ROC similar results in terms of classes. 

%\begin{figure*}[]
%  \centering

%  \subfigure[]{\includegraphics[height=6cm]{images/gcn_ps_ds2.png}}\quad  
%  \subfigure[]{\includegraphics[height=6cm]{images/gat_pr_ds2.png}}\quad
%  \subfigure[]{\includegraphics[height=6cm]{images/gcn_roc_ds2.png}}\quad
%  \subfigure[]{\includegraphics[height=6cm]{images/gat_roc_ds2.png}}\quad

%  \caption{Evaluation of binary termination estimation for Dataset 2 (DS2) using comparing regular GCN-layers (a \& c) with GAT-layers (b \& d). The estimation includes the precision-recall curves (a \& b) and receiver operating characteristic curves (c \& d). Non-terminating programs are assigned the class label 0 and terminating programs are assigned the label 1.}
  %A superior performance for graph attention layers is evident by higher AUC values. Additionally, the difference in precision-recall per class shows a higher AUC for class 0 (non-terminating programs). }
%  \label{fig:binResults}
%\end{figure*}

\subsection{Semantic segmentation of nodes causing nontermination}

In this section, we describe our experiments for identifying the outermost nonterminating %loop using semantic segmentation.
node using semantic segmentation. 

\subsubsection{Segmentation datasets}\label{sec:datasets-segmentation}

Several reasons prevent us from reusing the same benchmarks from DS-SV-COMP and DS-TERM-COMP for semantic segmentation. Firstly, they only contain a program-level label denoting whether the program terminates. For semantic segmentation, we require node-level annotations that indicate whether the corresponding AST node is likely to cause nontermination or not. 
Secondly, it is often very difficult to determine the cause for nontermination in existing benchmarks, as it isn't labeled. 

For this reason, we generate two datasets that consist only of nonterminating samples (as we would only attempt to find the cause for nontermination for those programs that were deemed nonterminating by our classifier). 
As explained in Section~\ref{sec:discussion-debugging}, we restrict our annotations to only identify the outermost nonterminating loop in the program. Note that if there are several distinct, outermost, nonterminating loops, only one of them will be labeled as causing nontermination. The reason for this is the fact that we use fuzz testing to determine the culpable loop, and we stop the fuzzing
once we found the first nonterminating scenario. We use label 0 to indicate that the corresponding AST node has no relation to nontermination, and label 1 to indicate a high likelihood of causing nontermination.

The two datasets are described in the last two rows of Table~\ref{tab:datasets}. The first, DS-SegPy1 contains Python programs, whereas the second, DS-Seg-C, contains C programs. Each dataset is composed of 180 programs for training and 50 for testing.
%The second dataset DS-SegC contains non-terminating C programs and serves as a reference to DS-Seg-Py1.
%We use both The different conversion from a sample program in C and Python to an AST might reflect in a different segmentation performance.
The program sample generator for custom data is probabilistic and generates programs that contain from two to five loops that can be nested. 

\subsubsection{Training}

To improve and continue training beyond binary cross-entropy convergence, we deployed a focal loss \cite{LossFunctions, FocalLoss}, an extension of cross-entropy that down-weights simple samples and gives additional weight to hard negatives.

To achieve optimal results, the experiments required modifying the learning rate, optimizer, network structure, and the number of features compared to those used for classification. The final configuration features an Adam optimizer with a learning rate of 0.001. In our experiments, we found that a number of 4 layers was sufficient for both the GCN and GAT architectures, respectively. While the GCN and GAT architectures used for semantic segmentation are respectively similar to those used for classification, they do not contain a node mean function and linear layers. The last GCN/GAT layer projects to a one-dimensional feature indicating the confidence that a node is causing nontermination. 

Similar to the experiments for classification, we perform a total of ten training sessions until convergence of the test metrics. 
%As the model capacity depends on the complexity of program dependency, the number of layers corresponds to the node's distance necessary to associate features necessary to estimate non-termination cause. For larger programs, a higher number of layers is required.

\subsubsection{Results and evaluation}

%[paragraph zero: Metrics]
The metrics used for evaluation are Jaccard loss based on intersection over union (IoU),  the related Dice-Coefficient, and node-wise accuracy. A comprehensive overview of these metrics is provided by \cite{evaluationMetrics}. Similar to image segmentation where the pixel-wise accuracy is determined, here the node-wise distinction of true-positives, true-negatives, false-positives, and false-negatives for the node prediction compared to the annotated node ground truth is used. 

We calculate the mean metrics for an average of 10 models that were trained to a convergence of validation scores for each experimental configuration. In the evaluation, we provide mean values with standard deviation. 

%[first paragraph (no diff btw positive and negative classes)] =====================

From the results for semantic segmentation (detailed in Table \ref{tab:segmentation}) we can infer several observations. With mean values of more than 0.84 for Dice and IoU and a node accuracy of 0.81 in all experimental validation sets, the ability to generalize for unknown programs is high. The low standard deviation indicates a robust result for the state of convergence of validation metrics.

%But the ability to generalize for larger programs and largely unknown types of programs cannot be estimated well. 
%We assume that for larger programs with a higher variable dependency the complexity of prediction will increase. It is difficult to estimate if a variable dependency for larger examples can be learned by the graph network. 

%[second paragraph GCN vs. GAT] =========================================
For dataset Seg-Py1 the mean Dice-Coefficient is higher for the GAT architecture by 3.3 percent than for GCN. Similarly, node accuracy is better by 4.7 percent with only a decrease of 0.8 percent for the IoU. This indicates an improvement in segmentation performance for GAT compared to GCN. Similarly, for C programs using dataset Seg-C, all metrics improved significantly with a 5 percent better performance for Dice-Coefficient and node accuracy and an increase of 12 percent for IoU. 
From the comparative better performance of GAT-layers for node-wise segmentation, we infer that the use of self-attention mechanisms enables the network to weigh more relevant node transitions and therefore achieve better performance. Higher attention is assigned to edges with a high likelihood of causing nontermination.

\subsection{Comparison to recurrent algorithms}

In Section~\ref{sec:intro}, we discussed the suitability of GNNs for analyzing programs given the inherent graph structure of programs.
The objective of the current section is to justify this statement by evaluating whether GNNs are better at estimating program termination than recurrent algorithms. Note that the latter
are based on a sequential program representation.

For this purpose, we create an additional set of experiments for the binary classification of program termination. 
To make the comparison fair, the experimental configuration ensures that the number of learnable parameters of all recurrent algorithms is higher than the learnable parameters of the GNNs they were compared against.
We use the PyTorch implementations for RNNs \cite{rnn}, Long short-term memory (LSTM) \cite{lstm} and Gated recurrent unit's (GRU) \cite{gru}.

For this experiment, we only use the DS2 dataset containing Python programs. Conversely to the previous experiments, for the recurrent algorithms, the programs are not converted into an AST but are directly encoded into a one-hot encoding based on a dictionary of unique instructions.  

For a significant evaluation, we perform a total of ten training sessions until convergence of the test metrics.

\subsubsection{Results}

%[paragraph zero: metrics]

Similar to the previous classification experiments we record the Precision-Recall (PR) and Receiver Operating Characteristic (ROC) and their respective AUC. 

%[paragraph one: compare general]
We summarize our experimental results in Table~\ref{tab:recurrent}. Notably, graph-based methods outperform recurrent approaches by more than 5 percent for the best recurrent ROC-AUC and PR-AUC metrics. Intuitively, we infer that the graph representation of programs is a better fit for termination estimation than the sequential representation. Furthermore,  the understanding of the context within the recurrent networks is limited by the model capacity and by the natural constraints that don't allow for large sequences of data to be processed.

\begin{figure}[]
  \centering
  
  \includegraphics[height=6.5cm]{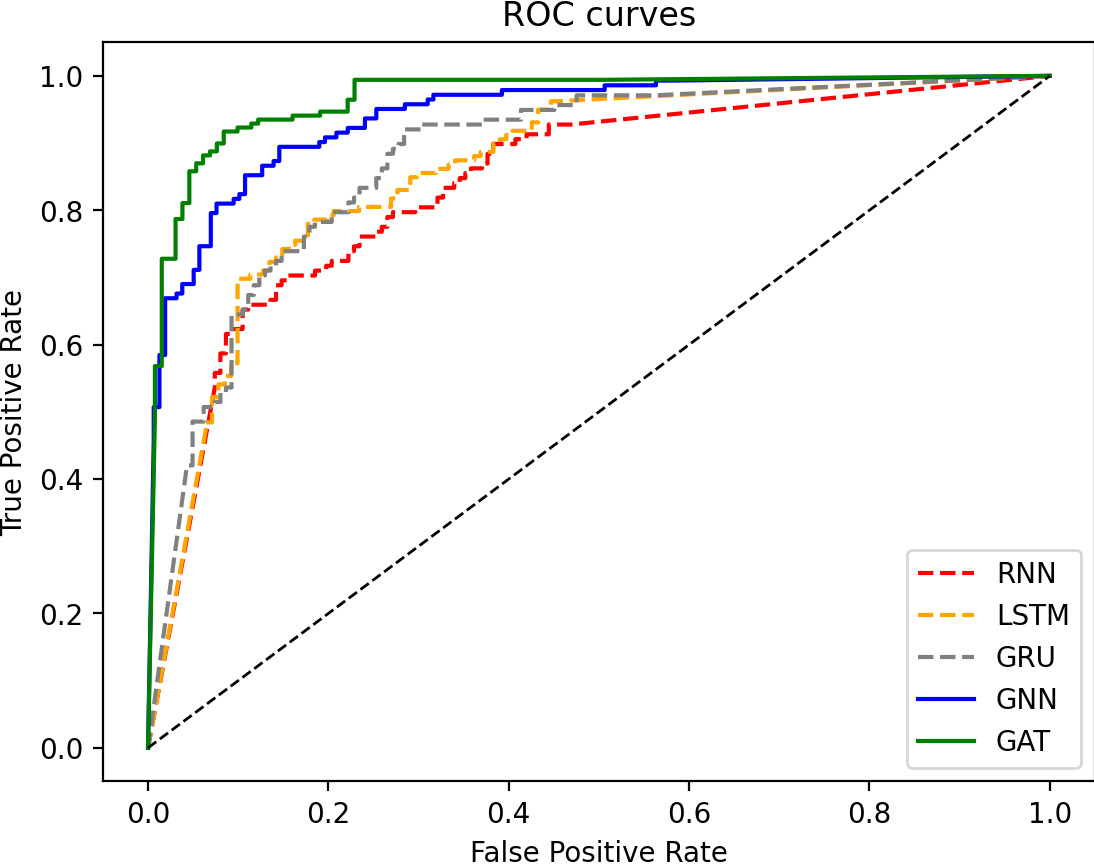}
  %\subfigure[]{\includegraphics[height=6.2cm]{images/rnngnnPR.png}}\quad
  %\subfigure[]{\includegraphics[height=6.2cm]{images/rnngnnROC.png}}\quad
  
  \caption{Binary classification for termination in recurrent and graph neural networks. The graph neural network based approaches perform significantly better than classical recurrent approaches.}% (a) Precision-Recall curve (b) ROC-Curve. }
  \label{ResultsRecurrent}
\end{figure}

Additional insight into the comparison of the classification results can be gained by examining Figure~\ref{ResultsRecurrent}, where it can be observed that the GNN and GAT-based architectures perform better than the recurrent approaches. 
%\subsection{Practical observations}

%recursion
%In recursive function calls the depth of recursion is usually checked by a compiler, where issues with endless recursion are usually avoided. Where mostly random variables are not considered as program input it is clear that the termination of a program might often depend on a random variable connected to an 'if'-statement for example. We, therefore, considered the results of random functions as part of the input set as well to stay within the definition of the halting problem. The sample is true for dynamic user inputs in run time.

%Inference
%Since the determination of halting for a program can be decided in the majority of programs by the first-order node with two graph convolutional layers we suggest a recursive process of several classification iterations for inference. After each convolution, a classification can be executed to determine a confidence score for halting. If the score is below a certain threshold another convolution is executed etc. With previously defined layer-depending thresholds and a maximum (iteration) depth inference is based on intrinsic estimation of reliability. 

\begin{table*} 
  \centering 
  \begin{tabular}{l c c c c c c c c c}   
  & \multicolumn{4}{c}{\textbf{Precision-Recall}}& \multicolumn{1}{c}{\textbf{ROC}} \\ 
  \cmidrule(l){3-5} 
  \cmidrule(l){6-8} 
  Approach & Dataset &  mAP & AUC negative & AUC positive & mAP &  AUC neg. & AUC pos.\\ 
  \midrule   
  \textbf{GCN} & DS-SV-COMP &  {\color{blue} 0.83} & 0.94 & 0.76 & {\color{blue} 0.87} &  0.88 & 0.88\\   
  & DS-TERM-COMP &  {\color{blue} 0.87} & 0.92 & 0.84 & {\color{blue} 0.90} &  0.89 & 0.89\\
  & DS1 &  {\color{blue} 0.92} & {\bf 0.94} & 0.90 & {\bf {\color{blue} 0.93}} &  0.92 & 0.92\\
  & DS2 &  {\color{blue} 0.87} & 0.91 & 0.88 & {\color{blue} 0.91} &  0.91 & 0.91\\   
  \midrule
  \textbf{GAT} & DS-SV-COMP & {\bf {\color{blue} 0.9}} & {\bf 0.96} & 0.86 & {\bf {\color{blue} 0.93}} &  {\bf 0.93} & {\bf 0.93}\\   
  & DS-TERM-COMP  & {\bf {\color{blue} 0.91}} & {\bf 0.93} & 0.89 & {\bf {\color{blue} 0.92}} & {\bf 0.91} & {\bf 0.91}\\   
  & DS1 & {\bf {\color{blue} 0.93}} & 0.91 & 0.74 & {\bf {\color{blue} 0.93}} &  {\bf 0.94} & 0.93\\   
  & DS2 &  {\bf {\color{blue} 0.90}} & {\bf 0.94} & 0.88 & {\bf {\color{blue} 0.92}} & {\bf 0.93} & {\bf 0.93}\\   
  \bottomrule 
  \end{tabular}
  \caption{Experimental results for the binary classification of program termination using different network architectures and datasets. ROC and Precision-Recall are recorded after a test accuracy convergence. } 
  \label{tab:clasResult} 
  \end{table*}

  \begin{table*} 
    \centering 
    \begin{tabular}{l c c c c c c c c c} 
    & & & \multicolumn{2}{c}{\textbf{Dice-Coefficient}}& \multicolumn{2}{c}{\textbf{Jaccard Index (IoU)}} & \multicolumn{2}{c}{\textbf{Node Accuracy}}\\ 
    \cmidrule(l){4-5} 
    \cmidrule(l){6-7} 
    \cmidrule(l){8-9} 
    \textbf{Approach} & Dataset & Language & value & $\sigma$ & value & $\sigma$ & value & $\sigma$ \\ 
    \midrule   
    \textbf{GCN} & Seg-Py1 & Python & 0.843 & 0.021 & 0.856 & 0.029 & 0.810 & 0.026 \\   
    
    & Seg-C & C &0.896 & 0.031 & 0.812 & 0.029 & 0.892 & 0.034 \\   

    \midrule
    \textbf{GAT} & Seg-Py1 & Python & 0.876 & 0.020 & 0.848 & 0.024 & 0.857 & 0.023\\        
     
     & Seg-C & C &0.947 & 0.012 & 0.943 & 0.024 & 0.944 & 0.013 \\   
    \bottomrule 
    \end{tabular}
    \caption{Experimental results for the semantic segmentation of termination using different network architectures and datasets. 
    } 
    \label{tab:segmentation} 
    \end{table*}

    \begin{table} 
      \centering 
      \begin{tabular}{l c c c c c c c} 

      \textbf{Approach} & Network & ROC-AUC & PR-AUC \\ %& Precision & Recall \\ 
      \midrule   
      \textbf{recurrent} & RNN \cite{rnn} & 0.78 & 0.77 \\ %& 0.71 & 0.73 \\   
      & LSTM \cite{lstm}  & 0.88 & 0.87 \\ %&0.76 & 0.84 \\       
      & GRU \cite{gru}  & 0.88 & 0.88 \\ %& 0.83 & 0.81\\       
      \midrule
      \textbf{graph based} & GCN (custom) & 0.94 & 0.94\\ % & 0.88 & 0.85\\       
      & GAT (custom)  & \textbf{0.96} & \textbf{0.96} \\ % &  \textbf{0.91} & \textbf{0.91}\\       
      \bottomrule 
      \end{tabular}
      \caption{Experimental comparison of binary classification using RNNs vs. GNNs. 
      } 
      \label{tab:recurrent} 
      \end{table}

\section{Related Work}

\subsection{Termination analysis} 

Termination analysis has been studied for a long time. While the majority of the approaches make use of symbolic methods such as loop summarisation~\cite{DBLP:journals/toplas/ChenDKSW18,DBLP:conf/pldi/0001K21}, program synthesis~\cite{DBLP:conf/cav/FedyukovichZG18,DBLP:conf/esop/DavidKL15}, quantifier elimination~\cite{DBLP:conf/ccs/Le00S18}, abstract interpretation~\cite{DBLP:conf/popl/CousotC12}, there were recent attempts to employ machine learning techniques.

For instance, Neural Termination Analysis \cite{NeuralTermination}, trains neural networks as ranking functions, i.e. monotone maps from the program's state space to well-ordered sets.

This method uses sampled executions to learn ranking functions, which are then
verified with an SMT solver.
Several limitations exist for this approach, e.g. loops with a limited number of iterations do not provide enough data to learn a ranking function,
the verification of the ranking function may require additional loop invariants to succeed. Moreover, this work has only been evaluated on terminating programs with a maximum of two nested loops.
Calude and Dumitrescu proposed a probabilistic algorithm for the Halting problem based on running times, where they define a class of computable probability distributions on the set of halting programs~\cite{HaltingProbablistic}. 
In~\cite{DBLP:conf/cav/AbateGR20}, Abate et al. present the first machine learning approach to the termination analysis of probabilistic programs, where they use a neural network to fit a ranking supermartingales over execution traces and then verify it over the source code with an SMT solver.

As opposed to these existing works, we do not attempt to provide strong guarantees about
the termination decision. Instead, our objective is to provide an estimation of a program’s termination behavior, as well as localizing the likely cause of nontermination (when
applicable) that a programmer can use for debugging purposes.
%% Instead of trying to prove that the network's decision is correct, we are interested in gaining insights into the reasons behind its decision.
For this purpose, we employ the attention mechanism to identify those nodes
relevant for the termination estimation. Moreover, for programs classified as nonterminating, we use semantic segmentation to distinguish the outermost loop causing the infinite execution.  

\subsection{Graph Neural Networks}

The original approach to GNNs as presented by Kipf and Welling \cite{GraphConvLayer} used the sum of normalized neighbor embeddings as aggregation in a self-loop. With a Multi-Layer-Perceptron as an aggregator, Zaheer et al. \cite{GraphMLP} presented an approach that propagates states through a trainable MLP. With the development of advanced attention networks, the approach of Velickovic et al. \cite{GraphAttention} focused on attention weights that allow to prioritize the influence of features based on self-learned attention. For heterogeneous graphs with additional edge features, Relational Graph Convolution Networks were introduced by Schlichtkrull et. al. \cite{RGCN} to enable link prediction and entity classification, allowing the recovery of missing entity attributes for high-dimensional knowledge graphs. 

\section{Conclusions}
We proposed a technique for estimating the termination behavior of programs using a GNN.
We also devised a GAT architecture that uses a self-attention mechanism to allow the visualization of nodes relevant for the termination decision. Finally, for nonterminating programs, we constructed a GAT for the semantic segmentation of those nodes likely responsible for nontermination.

Experimental evaluation of our algorithm confirms a superior performance in termination estimation with a high ability to generalize for unknown programs.

%% \begin{acks}
%% Cambridge Royal Society
%% \end{acks}

\nocite{*}

\bibliographystyle{ACM-Reference-Format}
\bibliography{gnn}

\end{document}